%==================================================================================================%
% File Name:               Dark_Metric.tex
% Generation:              5.0
% Last Modification Date:  29/06/2008
%==================================================================================================%

\def\makeatletter{\catcode`\@11\relax}
\makeatletter

\font\rmAddress=cmti10 scaled 900

\font\bfAbs=cmbx10
\font\Abs=cmr10
\font\bfTitle=cmbx12 scaled 1200
\font\bfSection=cmbx12 scaled 1000
\font\bfSubsection=cmbx12 scaled 900

\font\bibliography=cmsl8

\def\SecTag#1{#1}

\def\list#1{{\relind=1.5\parindent#1}}

\def\vvv{{\vskip 3\baselineskip}}
\def\v{{\vskip \baselineskip}}
\def\vv{{\vskip 2\baselineskip}}

\newskip\vskipamount \vskipamount=\baselineskip
\newskip\vvskipamount \vvskipamount=2\baselineskip

\def\vb{\ifdim\lastskip<\vskipamount\removelastskip\penalty-100\v\fi}
\def\vvb{\ifdim\lastskip<\vvskipamount\removelastskip\penalty-200\vv\fi}
\def\vvvb{\ifdim\lastskip<\vvvskipamount\removelastskip\penalty-300\vvv\fi}
\def\vvnb{\ifdim\lastskip<\vvskipamount\removelastskip\penalty1000\vv\fi}

\def\bib#1{[\nobreak{{#1\nobreak\hskip.1em\nobreak}}]}
\def\bib#1#2{\item{[#1]}{#2}}

\def\bibref#1{[\nobreak{\bf #1}]}

\def\noin{\noindent\hskip\absind}
\newdimen\relind
          \relind=\parindent
\newdimen\absind
   \absind=0 pt

\newskip\vvvskipamount \vvvskipamount=3\baselineskip
\def\vvvnb{\ifdim\lastskip<\vvvskipamount\removelastskip\penalty1000\vvv\fi}
\def\halfv{{\vskip.5\baselineskip}}
\def\title#1{{\centerline{\bfTitle #1}\par\vvvnb}}
\long\def\authors#1{\centerline{#1}\halfv}

\long\def\address#1#2{\item{#1}{\rmAddress#2}}

\def\section#1#2{\vvvb\noin{\bfSection #2.~#1}\vvb}
\def\subsection#1#2{\vvb\noin{\bfSubsection #2.~#1}\vb}
\def\subsubsection#1#2{\vvb\noin{\bf #2.~#1}\vb}
\long\def\abstract#1{\vvb{\leftskip 2cm
 \rightskip1cm\baselineskip14pt{{\noindent \bfAbs Abstract:} \Abs
 #1}\vvv}
}

\def\Narrower{\advance\leftskip by\parindent}

\long\def\date#1#2{\vb\noin{\bf #1}\par\Narrower{\noin #2}\par\advance \leftskip by -\parindent\vb}

\newdimen\z@ \z@=0pt
\def\m@th{\mathsurround\z@}

\catcode`\|=11

\let\= = \equiv
\def\_#1{_{\!#1}}
\def\||#1{{\left\Vert#1\right\Vert^{\kern-0.5em\phantom{1}}_{\kern-0.5em\phantom{1}}}}
\def\|#1{{\left\vert#1\right\vert^{\kern-0.5em\phantom{1}}_{\kern-0.5em\phantom{1}}}}
\def\(#1{{\left(#1\right)^{\kern-0.5em\phantom{1}}_{\kern-0.5em\phantom{1}}}}
\def\[#1{{\left[#1\right]^{\kern-0.5em\phantom{1}}_{\kern-0.5em\phantom{1}}}}
\def\{#1{{\left\lbrace#1\rbrace^{\kern-0.5em\phantom{1}}_{\kern-0.5em\phantom{1}}}}
\let\al=\alpha

\def\calL{{\cal L}}
\def\cases#1{\def\\{\cr}
          \left\lbrace\>\vcenter{\normalbaselines\m@th
 \ialign{$##\hfil$&&\quad$##\hfil$\crcr#1\crcr}}\right.^{\kern-0.5em\phantom{1}}_{\kern-0.5em\phantom{1}}}
\def\christ#1#2{\left\lbrace{}^{#1}_{#2}\rbrace}
\def\d{{\rm d}}
\def\de{\partial}
\def\dede#1#2{{\de #1\over\de #2}}

\def\deldel#1#2{\frac{\delta #1}{\delta #2}}

\def\deq{:\=}

\def\fnote#1#2{\relind=1.7\parindent
                ${}^{#1}$\insert\footins{
                                  \interlinepenalty100
                  \splittopskip=10pt plus 1pt minus 1pt
                  \floatingpenalty=20000
                  \smallskip
                  \absind=0pt
                  \item{#1}{#2}}}

\def\frac#1#2{{#1\over#2}}

\mathchardef\gamma="710D %§ Questa deve precedere tutte le macro che utilizzano la c.s. \gamma
\mathchardef\Gamma="7100 %§ Questa deve precedere tutte le macro che utilizzano la c.s. \Gamma

\let\Ga=\Gamma

\mathchardef\lambda="7115 %ß
\mathchardef\Lambda="7103 %ß
\let\la=\lambda
\let\La=\Lambda
\let\mi=\mu
\def\mini{{\mu\nu}}
\let\na=\nabla
\def\Na#1{\buildrel #1\over\na}

\let\ni=\nu

\def\R{{\rm R}}
\def\Ric{{\rm Ric}}
\def\rbrace{\right\}}

\def\sqr#1#2{{\vcenter{\vbox{\hrule height.#2pt
 \hbox{\vrule width.#2pt height#1pt \kern#1pt
 \vrule width.#2pt}
 \hrule height.#2pt}}}
}
\def\square{\mathchoice\sqr74\sqr74\sqr{2.1}3\sqr{1.5}3\>}

\def\Tag#1{\ifmmode\eqno{(#1)}\hbox to\rightskip{\null}\else(#1)\fi}
\def\Ref#1{(#1)}

\def\ie{{\it i.e.}}
\def\eg{{\it e.g.}}

\title{From Dark Energy \& Dark Matter to Dark Metric}

\halfv
\authors{S. Capozziello${}^*$, M. De
Laurentis${}^{\flat}$, M. Francaviglia${}^{\dagger,\ddagger}$, S.
Mercadante${}^{\dagger}$, }

\address{${}^*$}{Dipartimento di Scienze Fisiche, Universit\`a di Napoli "Federico II", and INFN
Sez.\ di Napoli, Compl.\ Univ.\ di Monte S.~Angelo, Edificio G, Via Cinthia, I-80126~-~Napoli,
Italy}

\address{${}^\flat$}{Dipartimento di Fisica, Politecnico di Torino and INFN Sez. di
Torino, Corso Duca degli Abruzzi 24, I-10129~-~Torino, Italy}

\address{${}^\dagger$}{Dipartimento di Matematica, Universit\`a di Torino, and INFN Sez.  di Torino,
Via Carlo Alberto 10, I-10123~-~Torino, Italy}

\address{${}^\ddagger$}{Laboratorio per la Comunicazione Scientifica dell'Universit\`a della
Calabria, Ponte Bucci, Cubo 30b, I-87036~-~Arcavacata di Rende (CS), Italy.}

\abstract{It is nowadays clear that General Relativity cannot be the definitive theory of
Gravitation due to several shortcomings that come out both from theoretical and experimental
viewpoints.  At large scales (astrophysical and cosmological) the attempts to match it with the
latest observational data lead to invoke Dark Energy and Dark Matter as the bulk components of the
cosmic fluid.  Since no final evidence, at fundamental level, exists for such ingredients, it is
clear that General Relativity presents shortcomings at infrared scales.  On the other hand, the
attempts to formulate more general theories than Einstein's one give rise to mathematical
difficulties that need workarounds that, in turn, generate problems from the interpretative
viewpoint.  We present here a completely new approach to the mathematical objects in terms of which
a theory of Gravitation may be written in a first-order ({\it \`a la\/} Palatini) formalism, and
introduce the concept of {\bf Dark Metric} which could completely bypass the introduction of
disturbing concepts as Dark Energy and Dark Matter.}

%--------------------------------------------------------------------------------------------------%
\section{Introduction}
         {\SecTag{1}}

Einstein General Relativity ({\bf GR}) is a self-consistent theory that dynamically describes space,
time and matter under the same standard.  The result is a deep and beautiful scheme that, starting
from some first principles, is capable of explaining a huge number of gravitational phenomena,
ranging from laboratory up to cosmological scales.  Its predictions are well tested at Solar System
scales and give rise to a comprehensive cosmological model that agrees with the Standard Model of
particles, with the recession of galaxies, with the cosmic nucleosynthesis and so on.

Despite these good results, the recent advent of the so-called {\it Precision Cosmology} and
several tests coming from the Solar System outskirts (\eg\ the Pioneer Anomaly) entail that the
self-consistent scheme of GR seems to disagree with an increasingly high number of observational
data, as \eg\ those coming from IA-type Supernovae, used as standard candles, large scale structure
ranging from galaxies up to galaxy superclusters and so on.  Furthermore, being not renormalizable,
GR fails to be quantized in any ``classical'' way (see \bibref{1}).  In other words, it seems then,
from ultraviolet up to infrared scales, that GR is not and cannot be the definitive theory of
Gravitation also if it successfully addresses a wide range of phenomena.

Many attempts have been therefore made  both to recover the
validity of GR at all scales, on the one hand, and to produce
theories that suitably generalize Einstein's one, on the other
hand.

Besides, in order to interpret a large number of recent
observational data inside the paradigm of GR, the introduction of
Dark Matter ({\bf DM}) and Dark Energy ({\bf DE}) has seemed to be
necessary: the price of preserving the {\sl simplicity\/} of the
Hilbert Lagrangian has been, however, the introduction of rather
odd-behaving physical entities which, up to now, have not been
revealed by any experiment at fundamental scales. In other words,
we are observing the large scale effects of missing matter (DM)
and the accelerating behaviour of the Hubble flow (DE) but no final
evidence of these ingredients exists, if we want to deal with them
under the standard of  quantum particles or quantum fields. In
Section~3, we shall argue whether, after all, it is really
preferable the use of the {\sl simplest\/} Lagrangian.

An opposite approach resides in the  so-called {\bf Non-Linear
Theories of Gravitation} ({\bf NLTGs}, see later), that have been
also investigated by many authors, also in connection with {\bf
Scalar-Tensor Theories} ({\bf STTs}, see later).  In this case, no
ill-defined ingredients have to be required, at the price of big
mathematical complications.  None of the many efforts made up to
now to solve this problem (see later) seems to be satisfactory
from an interpretative viewpoint.
\vb

What we shortly present here is a completely  new approach to the
mathematical objects in terms of which a theory of Gravitation may
be written, whereby Gravity is encoded from the very beginning in
a (symmetric) linear connection in SpaceTime.  At the end, we
shall conclude that, although the gravitational field is a linear
connection, the fundamental field of Gravity turns out {\it a
posteriori\/} to be still a metric,  but not the ``obvious'' one
given from the very beginning (which we shall therefore call {\bf
Apparent Metric}).  Rather we shall show the relevance of another
metric, ensuing from gravitational dynamics, that we shall call
{\bf Dark Metric} since we claim it being a possible source of the
{\sl apparently\/} ``Dark Side'' of our Universe which reveals itself, at
large scales, as missing matter (in clustered structures) and
accelerating behaviour (in the Hubble fluid).
\vb

To complete our program, we need first to  recall some facts
regarding different (relativistic) theories of Gravitation.  This
will not be an historical {\it compendium}, but rather a collection
of speculative hints useful to our aims.

%--------------------------------------------------------------------------------------------------%
\section{A critical excursus}
         {\SecTag{2}}

The theory of Special Relativity ({\bf SR}), published by A.\
Einstein in 1905  \bibref{2}, was aimed to reconcile Mechanics
with Electromagnetism, but leaved out matter and Gravitation
(Minkowski SpaceTime is rigorously empty and flat). Then Einstein
devoted  more than ten years (1905--1915/1916) to develop a theory
of Gravitation based on the following requirements (see
\bibref{3}): {\bf principle of equivalence} (Gravity and Inertia
are indistinguishable; there exist observers in free fall, \ie\
inertial motion under gravitational pull); {\bf principle of
relativity} (SR holds pointwise; the structure of SpaceTime is
pointwise Minkowskian); {\bf principle of general covariance}
(``democracy'' in Physics); {\bf principle of causality} (all
physical phenomena propagate respecting the light-cones).
Einstein, who was also deeply influenced by Riemann's teachings
about the link between matter and curvature, decided then to
describe Gravity by means of a (dynamic) SpaceTime $M$ endowed
with a dynamic Lorentzian metric $g$.  This appeared to be a good
choice for a number of reasons: a metric is the right tool to
define measurements (rods \& clocks); the geodesics of a metric
are good mathematical objects to describe the free fall; a
Lorentzian manifold is pointwise Minkowskian, is suitable to be
the domain of tensor fields, is compatible with a light-cones
structure.  And, after all, at that time, there was no other
geometrical field Einstein could use to define the curvature of a
differentiable manifold!

Following this way, Einstein deduced his famous equations:
$$
G_\mini\deq\R_\mini-\frac{1}{2}\>\R(g)\>g_\mini=8\pi\>G\>T_\mini~.
\Tag{1}
$$
A linear concomitant of the Riemann curvature tensor of $g$, nowadays called the Einstein tensor
$G_\mini$, equals here the stress-energy tensor $T_\mini\deq\deldel{L_{\rm mat}}{g_\mini}$ that
reflects the properties of matter.  Here $\R_\mini$ is the Ricci tensor of the metric $g$ and
$\R(g)\deq g^\mini\>\R_\mini$ is the scalar curvature of the metric, while $L_{\rm mat}\=\calL_{\rm
mat}\>\d s$ is the matter Lagrangian and $G$ is the gravitational coupling constant.

In other words, the distribution of matter influences Gravity through 10 second-order field
equations.  Their structure, in a sense and {\it mutatis mutandis}, is the same as Newton second law
of Dynamics: no forces means geodesic motion, while the effects of sources are to produce curvature
(just in motion in the Newtonian case, where the Space and Time are separately fixed and immutable;
both in the structure of SpaceTime and in its motions in Einstein's case).

GR has been a success: it admits an elegant and very {\sl simple\/} Lagrangian formulation (the
Lagrangian is $L_{\rm H}\deq\R(g)\>\sqrt{g}\>\d s$ and it was first found by Hilbert in 1915) and
most of its predictions have soon been experimentally verified and these have remained valid for
many years after its introduction.  So there was no reason for Einstein to be unhappy with his
beautiful creation, at least for some time.

In GR, is $g$ the gravitational field?  Einstein knew that it is not, since $g$ is a tensor, while
the principle of equivalence holds true and implies that there exist frames in which the
gravitational field can be inertially switched off, while a tensor cannot be set to vanish at a
point in a frame, if it does not vanish at that point in all frames.  Free fall in GR is in fact
described by the geodesics of $(M, g)$
$$
\ddot x^\la+\christ{\>\la}{\mini}_g\>\dot x^\mi\>\dot x^\ni=0~.
\Tag{2}
$$
Einstein himself argued that  the right objects to represent
the gravitational field have to be the Christoffel symbols
$\christ{\>\la}{\mini}_g$; the metric $g$ is just the potential of
the gravitational field,  but being the Christoffel symbols
algorithmically constructed out of $g$, the metric remains the
fundamental variable: $g$ gives rise to the gravitational field,
to causality, to the principle of equivalence as well as to rods
\& clocks.
\vb

In 1919, working on the  theory of ``parallelism'' in manifolds,
Tullio Levi-Civita understood that parallelism and curvature are
non-metric features of space, but rather features of ``affine''
type, having to do with ``congruences of privileged lines'' (see
\bibref{4}).  Generalizing the case of the Christoffel symbols
$\christ{\>\la}{\mini}_g$ of a metric $g$, Levi-Civita introduced
the notion of {\bf linear connection} as the most general object
$\Ga^\la_\mini$ such that the equation of geodesics
$$
\ddot x^\la+\Ga^\la_\mini\>\dot x^\mi\>\dot x^\ni=0
\Tag{3}
$$
is generally covariant.  This revolutionary idea  (that stands in
fact at the heart of Non-Euclidean Geometries) has been
immediately captured by Einstein who, unfortunately, did not
further use it up to its real strength.  We shall come back later
on this topic, as this work is strongly based on it.

For now, let us just reconsider GR in  terms of the new concept
introduced by Levi-Civita.  In GR free-falls are described by
equation \Ref{2}.  So, in GR, there is  a connection $\Ga$, but it
is given from the very beginning as the Levi-Civita connection of
the metric $g$: we write $\Ga\=\Ga_{\rm LC}(g)$, \ie\ locally
$\Ga^\la_\mini=\christ{\>\la}{\mini}_g$.  This has been in fact
realized and used by Attilio Palatini in the same year to re-write
the derivation of Einstein equations by using step-by-step only
tensorial quantities depending on $g$ (see \bibref{5}).  Let us
notice that the relations $\Ga^\la_\mini=\christ{\>\la}{\mini}_g$
are not ``essential'' equations: they express a founding
assumption on SpaceTime structure.  The connection $\Ga_{\rm
LC}(g)$ has no independent dynamics.  Only $g$ has dynamics and
$\Ga_{\rm LC}(g)$ behaves accordingly.  Thus, as we already said,
the single object $g$ determines at the same time the causal
structure (light cones), the measurements (rods \& clocks) and the
free fall of test particles (geodesics).  SpaceTime is definitely
a couple: $(M,g)$.

Even if it was clear to Einstein that Gravity  induces ``freely
falling observers'' and that the principle of equivalence selects,
in fact, an object that cannot be a tensor, since it is capable of
being ``switched off'' and set to vanish at least at a point, he
was obliged to choose this object under the form of the linear connection
$\Ga_{\rm LC}(g)$, fully determined by the metric structure
itself. Einstein, for obvious reasons, was very satisfied of
having reduced all SpaceTime structure and Gravity into a single
geometrical object, namely the metric $g$.
\vb

Still, in 1919, Hermann Weyl tried (see \bibref{6}) to unify Gravity with Electromagnetism, using
for the first time a linear connection defined over SpaceTime, assumed as a dynamical field
non-trivially depending on a metric.  Weyl's idea, unfortunately, failed because of a wrong choice
of the Lagrangian and few more issues, but it generated however a keypoint: connections may have a
physically interesting dynamics.
\vb

Einstein  soon showed a great interest in Weyl's idea.  He too
began to play with connections, in the obsessed seek for the
``geometrically'' Unified Theory.  But he never arrived to
``dethronize'' $g$ in the description of the gravitational field.
Probably, at some moments, he was not so happy with the fact that
the gravitational field is not the fundamental object, but just a
by-product of the metric; however, he never really changed his
mind about the physical and mathematical role of $g$.

In 1925 Einstein constructed a theory that depends on a metric $g$ and a symmetric linear connection
$\Ga$, to be varied independently (the so-called {\bf Palatini method}, because of a
misunderstanding with W.\ Pauli; see \bibref{5}); he defined a Lagrangian theory in which the
gravitational Lagrangian is
$$
L_{\rm PE}\deq \R(g,\Ga)\>\sqrt{g}\>\d s~,
\Tag{4}
$$
with
$$
\R(g,\Ga)\deq g^\mini\>\R_\mini(\Ga,\de\Ga)
\Tag{5}
$$
the Òscalar curvatureÓ of both $g$ and $\Ga$, being $\R_\mini(\Ga,\de\Ga)$ the Ricci tensor of
$\Ga$.

There are now 10 + 40 independent variables  and the field
equations, in vacuum, are:
$$
\cases{
 \R_{(\mini)}-\frac{1}{2}\>\R(g,\Ga)\>g_\mini=0\\
 \Na{\Ga}_\al(\sqrt{g}\>g^\mini)=0}
\Tag{6}
$$
where $\R_{(\mini)}$ is the symmetric part of $\R_{\mini}(\Ga,\de\Ga)$ and $\Na{\Ga}$ denotes
covariant derivative with respect to $\Ga$.

If the dimension $m$ of SpaceTime is greater than $2$, the second field equation \Ref{6}$_2$
constrains the connection $\Ga$, which is {\it a priori\/} arbitrary, to coincide {\it a posteriori}
with the Levi-Civita connection of the metric $g$ (Levi-Civita theorem).  By substituting this
information into the first field equation \Ref{6}$_1$, the vacuum Einstein equation for $g$ is
eventually obtained.  In Palatini formalism, the metric $g$ determines rods \& clocks, while the
connection $\Ga$ determines the free fall, but since {\it a posteriori\/} the same result of GR is
found, Einstein soon ceased to show a real interest in this formalism.

Let us now make  a digression. The situation does not change if matter
is present through a matter Lagrangian $L_{\rm mat}$ (independent
of $\Ga$ but just depending on $g$ and other external matter
fields), that generates an energy-momentum tensor $T_\mini$ as
$T_\mini\deq\deldel{L_{\rm mat}}{g_\mini}$. If the total
Lagrangian is then assumed to be $L_{\rm tot}\deq L_{\rm
PE}+L_{\rm mat}$ field equation \Ref{6}$_1$ are replaced by
$$
\R_{(\mini)}-\frac{1}{2}\>\R(g,\Ga)\>g_\mini=8\pi\>G\>T_\mini
\Tag{7}
$$
and again \Ref{6}$_2$ implies,  {\it a posteriori}, that \Ref{7}
reduces eventually to Einstein equations \Ref{1}.

Let us also emphasize that  the dynamical coincidence between
$\Ga$ and the Levi-Civita connection of $g$ is entirely due to the
particular Lagrangian considered by Einstein, which is the {\sl
simplest},  but not the only possible one!  Furthermore, it seems
to us that Einstein did not fully recognize that the Palatini
method privileges  the affine structure with respect to the metric
structure.  In fact, the Einstein-Palatini Lagrangian is of order
zero in the metric, while it is first order in the connection; in
other words, it contains (first order) derivatives of $\Ga$ but no
derivatives of $g$.  The connection is the real dynamical field
while the metric $g$ has no dynamics, since it enters the
Lagrangian just as a ``Lagrange multiplier.'' This time is the
metric $g$ to gain a dynamical meaning from that of $\Ga$, that
plays the role of fundamental field.

Notice that, in this case (\ie\ in Palatini formalism), the
relations
$$
\Ga^\la_\mini=\christ{\>\la}{\mini}_g
\Tag{8}
$$
are field equations: the fact that $\Ga$ is the Levi-Civita
connection of $g$ is no longer an assumption {\it a priori\/} but
it is the outcome of field equations!

%--------------------------------------------------------------------------------------------------%
\section{Among the different Theories of Gravitation, should we really prefer the simplest (in the
sense of the one with the simplest Lagrangian)?}
{\SecTag{3}}

Let us shortly  criticize the final points that we stressed at
the end of Section~2.  We are now in a situation that was already
seen elsewhere in Mathematics and Physics, from which we should
try to learn something.  We refer to the onset of Non-Euclidean
Geometry.  Let us limit for simplicity ourselves to discuss plane
(2-dimensional) Geometry.  Notice first that Euclidean Geometry is
based on two fundamental and {\it a priori\/} distinct structures:
{\bf metricity} and {\bf linearity} (or, better, {\bf affinity}).
In modern language we say that the Euclidean plane carries the
({\it a priori\/} independent) structures of vector (or, better,
affine) space and of metric space.  One depends on the linear
group (or its affine extension) while the other on the (subgroup)
of orthogonal transformations.

In a plane, metricity  selects the circles (\ie\ the level sets of
the distance function) while affinity the congruence of all
straight lines and their parallelism properties.  This corresponds
to the well-known ``compass \& unmarked straightedge Geometry.''

One of the by-products  of the aforementioned works of
Levi-Civita, Weyl and Einstein is the demonstration that these two
structures are in fact separated.  One is not obtained directly
from the other.

Galilean and Newtonian  Physics are in fact strongly based on the
assumption that Space carries a Euclidean structure.  A famous
wording of Galilei (``Il Saggiatore'') tells that: ``{\sl to
understand the Universe we have to know the language in which it
is written and its characters.  The language is Mathematics, while
the characters are circles, triangles and other geometrical
figures.}'' Galilei states that Space has two structures: the
metric one (circles) and the linear one (triangles).

From a purely geometrical  viewpoint, these two structures stand
on an equal footing, but from a physical point of view the
situation is however different: the principle of Inertia selects,
in fact, the straight lines as the more fundamental structure,
while circles limit their role to the definition of space
distances!  The whole Euclidean Geometry is a subtle mix-up of
both structures, but the fundamental principle of Physics (the
First Law, \ie\ the principle of Inertia) privileges straight
lines, \ie\ (uniform) rectilinear motion, when forces are not
present. Forces (and metricity) will enter the game only through
the Second Law (\ie\ Newton law ${\bf F}=m\>{\bf a}$, where ${\bf
a}$ contains curvature, hence metric relations, and ${\bf F}$ is
the source that generates ${\bf a}$ as a deviation from uniform
and rectilinear motion).

These two ({\it a priori\/} distinct)  structures may nevertheless
be intertwined by the simplest variational principle: ``straight
lines are the shortest path between any two points.''  This
variational principle has two advantages and one disadvantage.
Advantages: (1) it connects the two fundamental ({\it a priori\/}
distinct) structures of Euclidean Geometry and of Newtonian
Physics; (2) it is the {\sl simplest}.  Disadvantage: our world is
not Euclidean!  (Physics  clearly shows that).

In Geometry, we are therefore forced by Nature to consider
variational principles different from the {\sl simplest} and to
introduce metric geometries that differ from the {\sl simple\/}
Euclidean one. This is exactly the way in which Non-Euclidean
Geometry arises (Gauss, 1836; Riemann, 1856). The metric defines a
distance function: circles are replaced by the level sets of it;
straight-lines are replaced by ``geodesics,'' \ie\ lines of
minimal length. The Euclidean plane is just an example where
geodesics are straight lines and ``circles'' are ordinary circles.

Why should be so strange that the same happens in Gravity?

Moreover, in the  light of Levi-Civita's work, we should also
notice that this variational assumption can be replaced by a
different one.  In a space endowed with a (symmetric) linear
connection $\Ga$, geodesics are no longer defined by a variational
prescription, but are defined as self-parallel lines (\ie, their
tangent vector is parallel-transported by the connection).  In
Euclidean Space, where a metric is given first, we could envisage
the following situation, that summarizes the content of the
already evoked Levi-Civita theorem: suppose a symmetric connection
is given in it; claim that its self-parallel curves are the
straight lines, \ie\ those lines that minimize distance; then the
connection is forced, {\it a posteriori}, to be the Levi-Civita
connection of the Euclidean metric.  The same will happen in any
Riemannian manifold $(M,g)$: if we give a symmetric $\Ga$ on it
and pretend that the self-parallel curves of $\Ga$ minimize the
Riemannian length, then $\Ga$ is forced to be the metric
connection $\Ga_{\rm LC}(g)$.  In a sense, there is a variational
prescription that relates the apparently un-related structures
implied by $g$ and $\Ga$ independently.  In a sense, this
variational prescription is the ``Palatini prescription'' based on
the Lagrangian $\R(g,\Ga)\deq g^\mini\>\R_\mini(\Ga,\de\Ga)$.

%--------------------------------------------------------------------------------------------------%
\section{So then?}{\SecTag{4}}

All that said, we believe we should first seriously reconsider
NLTGs, without being unsensitive with respect to the appeal of
{\sl simplicity}, in the spirit of Occam Razor. This is why we
begin to restrict ourselves to the first level of generalization
of GR, the so-called {\bf $f(\R)$-theories} of metric type (see
\eg\ \bibref{7} for a  review of the results concerning
these theories). Here $f$ denotes any ``reasonable'' function of
one-real variable. The Lagrangian is assumed to be
$$
L_{\rm NL}(g)\deq f(\R(g))\>\sqrt{g}\>\d s
\Tag{9}
$$
Of course, from $f(\R)$-theories, we know that GR is retrieved in,
and only in, the particular case $f(\R)\=\R$, \ie\ if and only if
the Lagrangian is linear in $\R$.
\vb

Let us recall here just a few  keypoints on metric
$f(\R)$-theories.

When treated in  the purely metric formalism, these theories are
mathematically much more complicated than GR. These theories do in
fact produce field equations that are of the fourth order in the
metric:
$$
f'(\R(g))\>R_\mini(g)-\frac{1}{2}\>f(R(g))\>g_\mini
  -\underbrace{(\na\_\mi\na\_\ni-g_\mini\square)\>f'(\R(g))}_{\hbox{$4^{\rm th}$ order term}}
  =8\pi\>G\>T\_\mini
\Tag{10}
$$
where $f'$ denotes the derivative of $f$ with  respect to its real
argument.  This is something that cannot be accepted if one
believes that physical laws should be governed by second order
equations. In \Ref{10} we see a second order part that resembles
Einstein tensor (and reduces identically to it if and only if
$f(\R)\=\R$, \ie\ if and only if $f'(\R)\=1$) and a fourth order
``curvature term'' (that again reduces to zero if and only if
$f(\R)\=\R$).
\vb

A first workaround that was suggested long ago  to this problem is
to push the 4th order part
$(\na\_\mi\na\_\ni-g_\mini\square)\>f'(\R(g))$ to the r.h.s. This
lets us to interpret it as an ``extra gravitational stress''
$T^{~{\rm curv}}\_\mini$ due to higher-order curvature effects,
much in the spirit of Riemann.  In any case, however, the fourth
order character of these equations makes them  unsuitable under
several aspects, so that they were eventually abandoned for long
time and only recently they have regained  interest (see
\bibref{7} and references therein).
\vb

A second way to tackle the problem  has been proposed in 1987 (see
\bibref{8}, based on earlier work by the same authors
\bibref{9}, together with the references quoted therein). Notice
that these are the first papers where the Legendre transformation
that introduces an extra scalar field has been ever considered in
literature (it has been later ``re-discovered'' by other authors),
so that its priority should always be appropriately quoted when
dealing with ``metric'' $f(\R)$-theories. This is a method {\it
\`a la\/} Hamilton, in which, whenever one has a non-linear
gravitational metric Lagrangian of the most general type $L_{\rm
GNL}(g)\deq f(g,\Ric(g))$, one defines a {\sl second\/} metric $p$
as
\fnote{1}
      {This idea corresponds to an Einstein's attempt, dating back to 1925, to construct a ``purely
       affine''
       theory  (see \bibref{10}), \ie\ a theory in which the only dynamical field is a linear
       connection.  In this theory no metric is given from the beginning, but since it is obviously
       necessary to have a metric, the problem arises of how to construct it out of a connection.
       Einstein firstly tried to define the metric as the symmetric part of the Ricci tensor
       constructed out of the connection.  But this idea could not work (unless for quadratic
       Lagrangians). A.\ Eddington then
       proposed a recipe analogous to \Ref{11}. In this way Einstein and Eddington obtained a theory that
       reproduces GR,
       without introducing anything new. That is why Einstein eventually abandoned it too.
	   About these purely affine
       theories see also \bibref{11} where J.\ Kijowski correctly pointed out that in the purely affine
       framework the prescription \Ref{11} of Einstein and Eddington is nothing but the assumption
       that the metric can be considered as a momentum canonically conjugated to the connection.}

$$
p_\mini\deq \dede{L_{\rm grav}}{\R_\mini}~.
\Tag{11}
$$
In this way  the second metric $p$, a canonically conjugated
momentum for $g$, is a function of $g$ together with its first and
second derivatives, since it is a function of $g$ and $\Ric(g)$,
the Ricci tensor of $g$.  Notice that this leads to two equations
of the second order in $g$ and $p$, as Hamilton method always
halves the order of the equations by doubling the variables.
Following this method in the simpler $f(\R)$ case one gets that
the ``auxiliary'' metric $p$ is related to the original one $g$ just by
a conformal transformation:
$$
p\=\phi\>g~,\qquad \phi\deq f'(\R(g))~.
\Tag{12}
$$
The Lagrangian equations \Ref{10} are then rewritable as a Hamiltonian system:
$$
\cases{
 {\rm Ein}(p)=T_{\rm mat}+T_{\rm KGnl}\\
 {\rm KGnl}(\phi)=0}
\Tag{13}
$$
where KGnl means non-linear Klein-Gordon (because of a potential depending on $f$; see \bibref{7},
\bibref{8}, \bibref{9} for details).  Rewritten in this form, the theory has now two variables: the
``auxiliary'' metric $p$ (or, equivalently, the original one $g$) and the scalar field $\phi$.  This
is why these theories belong to a wide sector of theories that are called {\bf Scalar-Tensor
Theories}.  For more details, and in particular for their application in Cosmology and Extragalactic
Astrophysics see, \eg, \bibref{7} and the references quoted therein.  See also \bibref{12}, where
equivalence properties between NLGTs and STTs are discussed in detail.

Notice that \bibref{8}, \bibref{9} and all subsequent literature
left in fact open a few fundamental problems: Who really are the
second metric $p$ and the scalar field $\phi$? How to interpret
them (the scalar field $\phi$ survives even in vacuum)?  And\dots
what about the original metric $g$?

Fortunately there is a third method to solve the problem.

%--------------------------------------------------------------------------------------------------%
\section{Palatini formalism revisited and the Dark Metric}{\SecTag{5}}

The third method  anticipated at the end of the previous Section
is the Palatini method applied to the case of $f(\R)$-theories.
Now SpaceTime is no longer a couple $(M,g)$ but rather a triple
$(M,g,\Ga)$, with $\Ga$ symmetric. The Lagrangian is assumed to be
the non-linear Palatini-Einstein Lagrangian
$$
L_{\rm NLPE}(g,\Ga)\deq f(\R(g,\Ga))\>\sqrt{g}\>\d s
\Tag{14}
$$
with $\R(g,\Ga)\deq g^\mini\>\R_\mini(\Ga,\de\Ga)$ and $f$
``reasonable.'' Field equations \Ref{6} are now replaced by the
following:
$$
\cases{
 f'(\R(g,\Ga))\>\R_{(\mini)}-\frac{1}{2}\>f(\R(g,\Ga))\>g_\mini=8\pi\>G\>T_\mini\\
 \Na{\Ga}_\al(f'(\R(g,\Ga))\sqrt{g}\>g^\mini)=0}
\Tag{15}
$$
that take into account a possible  Lagrangian of the type $L_{\rm
mat}\=L_{\rm mat}(g,\psi)$, with $\psi$ arbitrary fields coupled
to $g$ alone (and {\sl not\/} to $\Ga$).  Notice that \Ref{15}$_1$
reduces to \Ref{7} if and only if $f(\R)\=\R$.  Notice also that
the trace of equation \Ref{15}$_1$ gives
$$
\R(g,\Ga)\>f'(\R(g,\Ga))-\frac{m}{2}\>f(\R(g,\Ga))=8\pi\>G\>\tau
\Tag{16}
$$
being $\tau\deq g^\mini\>T_\mini$ the ``trace''  of the
energy-momentum tensor. This equation was called the {\bf master
equation} in \bibref{13} and was there at the basis of a subtle
discussion about ``universality'' of Einstein equations in non-linear
special cases (\ie, when $\tau\=0$). Notice also the analogy of
\Ref{16} with the trace of \Ref{10}, \ie
$$
f'(\R(g))\>\R(g)-\frac{m}{2}\>f(\R(g))+(m-1)\>\square f'(\R(g))=8\pi\>G\>\tau
\Tag{17}
$$
and notice that only in the peculiar  case $f(\R)\=\R$ they reduce
to the {\sl same\/} equation, namely \Ref{7}. In all other cases
\Ref{17} entails that non-linearity ($f'\not=1$) produces, in the
metric formalism, effects due to the scalar factor $f'(\R)$, \ie\
depending eventually on a scalar field tuned up by curvature.

Approaching $f(\R)$-theories {\it \`a la\/} Palatini, we may now follow \bibref{13} step-by-step and
make a number of considerations (well summarized also in the recent critical review \bibref{14}).
At the end of these considerations we may conclude that:
\list{%\relind=
 \item{(1)}{When (and only when) $f(\R(g,\Ga))\=\R(g,\Ga)$ we ``fully'' recover GR for the given
  metric $g$.}
 \item{(2)}{For a generic $f(\R(g,\Ga))$ and in presence of matter such that
  $\tau\deq g^\mini\>T_\mini\=0$ (and thence,
  in particular, in vacuum), the theory
  is still equivalent to GR for the given metric $g$ with a ``quantized'' cosmological constant
  $\La$ and a modified coupling constant. In this case, in fact, the master equation
  \Ref{16}  implies that the scalar curvature $\R(g,\Ga)$ has to be a suitable constant $\al$,
  possibly
  and usually not unique but always chosen in a set that depends on $f$, so that \Ref{15}$_2$ still
  implies \Ref{8} with an additional cosmological term $\La(\al)\>g_\mini$.}
 \item{(3)}{For a generic $f(\R(g,\Ga))$ but in presence of matter such that
  $\tau\deq g^\mini\>T_\mini\not\=0$ one can (implicitly)
  solve the master equation \Ref{16}, for $m\not=2$,  and obtain $\R(g,\Ga)$ as a function
  $\R(\tau)$ of the given trace $\tau$. Then, knowing $f$, one gets implicitly
  $f(\R(g,\Ga))=f(\tau)$ and $f'(\R(g,\Ga))=f'(\tau)$, so that equation \Ref{15}$_2$ tells us that
  $\Ga$ is forced to be the Levi-Civita connection $\Ga_{\rm LC}(h)$ of a new metric $h$,
  conformally related to the original one $g$ by the relation
  $$
  h_\mini\deq f'(\tau)\>g_\mini\=f'(\R(g,\Ga))\>g_\mini~.
  \Tag{18}
  $$
  Then, using again equation \Ref{15}$_1$, we see  that the theory could be still rewritable as in
  \Ref{12} in a purely metric setting, but with far less interpretative problems, as we can
  immediately show.  But, from a viewpoint ``{\it \`a la\/} Palatini}'' in a genuine sense the method
  has in fact generated a completely new perspective.  The remaining field equations \Ref{15}$_1$,
  in fact, are still equivalent to Einstein equations with matter (and cosmological constant)
  provided one changes the metric from $g$ to $h$!
}%

What conclusions may be drawn?

The most enlightening case is that of $f(\R)$ with generic matter
and $\tau\not=0$.  Here, in fact, the universality property (see
again \bibref{13}) does not hold in its strict form, but in an
interesting wider interpretation: the dynamics of the connection
$\Ga$ still forces $\Ga$ itself to be the Levi-Civita connection
of a metric, but not of the ``original'' metric $g$, which we
shall prefer to call the {\bf apparent metric} for a reason we
clarify in a moment.  Instead, the dynamics of $\Ga$ identifies a
new metric $h$, conformally related to the apparent one $g$, which
we call the {\bf Dark Metric}. The Dark Metric $h$, we claim, is
the true origin of the ``Dark Side of the Universe''!

The apparent metric $g$ is in fact the one by means of which we
perform measurements in our local laboratories.  In other words,
the metric $g$ is the one we have to use every day to construct
and read instruments (rods \& clocks).  This is why we like to
call it the ``apparent'' metric.  But we claim that the right
metric we have to use as the  fundamental object to describe
Gravity is, by obvious reasons, the Dark Metric, since it is the
one responsible for gravitational free fall through the
identification $\Ga\=\Ga_{\rm LC}(h)$. Notice, incidentally, that
photon world-lines and causality are not changed, since the
light-cones structure of $g$ and $h$ are the same by conformal
invariance.

In other words, in our laboratories we have to use the apparent
metric $g$, but in our Gravity theories the dark one $h$.  The
translation from one ``language'' to the other is nothing but the
conformal factor $f'(\R)$, which manifestly depends on the theory
and on the content in ordinary matter.  Let us also notice
explicitly that this in particular implies that if a certain
metric $h$ is expected to be a solution of a problem, from a
theoretical point of view, it is rather important to look for $h$
in experiments.  Testing our theories with $g$, in a sense, is
wrong, since it is the  conformally related metric $h$ to be
searched for instead!

% The fundamental field is the linear connection $\Ga$, that has a dynamics since it enters the
% Lagrangian together with its first derivatives.
% The metric $g$ is no longer a Lagrange multiplier, but still has no dynamics since it enters
% algebraically the Lagrangian.

%--------------------------------------------------------------------------------------------------%
\section{Subtle is the Lord\dots and malicious He is too!}{\SecTag{6}}

Quoting \bibref{15} we can say that ``subtle is the Lord,''  since
the apparent metric $g$ is not the right one in terms of which we
can describe the gravitational field.  But differently from
\bibref{15} we eventually conclude that ``malicious He is too,''
 because the Dark Metric is\dots
hidden, until one considers theories more complicated than GR,
namely $f(\R)$-theories.  The linear Hilbert Lagrangian, in fact,
hides completely the Dark Metric (thence the name we gave it),
since only in this specific and rather peculiar case (both with
and without matter) the Dark Metric $h$ coincides with the
apparent one $g$.  In all other cases $h\not=g$ and the (unknown)
conformal factor $\phi\=f'(\R(g,\Ga))\=f'(\tau)$ has to be
phenomenologically tested against observational data in order to
find which is the (class of) Lagrangian(s) $f$ that, given $\tau$
(\ie, given the ``visible matter''), allow one to interpret the
``supposed Dark Matter'' (and Energy) as a curvature effect
\bibref{16} due to the curvature of $\Ga$ (namely, Gravity) which, in
turn, pointwise changes rods \& clocks according to the conformal
curvature-dependent (and matter-dependent) rescaling of Eq. \Ref{18}.
In  forthcoming studies, we will give hints on how DE (accelerated
cosmic behaviour) and DM (clustered structures) could be
interpreted as Dark Metric effects. We shall also address the intriguing problem of the relations 
that certainly exist between our geometrical approach to the conformal factor $\phi=f'(\R)$ and 
the so-called ÒAWE hupotesisÓ of Alimi and coworkers (see, \eg, \bibref{17} and references quoted 
therein).

As a final remark, we would also like to stress that changing the metric from a given one, needed to
define measurements in the laboratory, to a different one which seems to be more fundamental for the
description of Physics, as it might be suggested by the theoretical structure or by phenomenology,
is not at all new in Physics.  Recall in fact that in Newtonian Physics, the metric of Space is
rigidly assumed to be the Euclidean one $e$; nevertheless, when dealing with (holonomically)
constrained systems of particles, this metric and the kinetic energy $T$ induce together another
metric $g$ in the appropriate configuration space, whereby dynamics is eventually governed by a
metric Lagrangian of the form $T-U$.  Analogously when dealing with continua (with or without
internal structure), where the Euclidean metric induces the deformation metric of the body.  In all
these cases Physical properties of the system are governed by another metric, which encompasses at
the same time the Euclidean one (\ie, the one that is used to perform measurements in the
laboratory) and the realm of physical phenomena.  We claim that here something analogous happens,
since the dark metric contains both the given ÒlocalÓ metric $g$ as well as the gravitational pull
of curvature as induced by the conformal factor that, in turn, is reminiscent of the linear
connection.

%--------------------------------------------------------------------------------------------------%
\section{Acknowledgements}{\SecTag{7}}

We would like to warmly thank Andrzej Borowiec,  Lorenzo Fatibene,
and Marcella Giulia Lorenzi for fruitful discussions.

\null

%--------------------------------------------------------------------------------------------------%
\section{Bibliography}{\SecTag{8}}

\bib{1}{R.\ Utiyama, B.\ S.\ DeWitt, J.~Math.\ Phys.\ {\bf 3} (1962), 608.}

\bib{2}{A.\ Einstein, {\sl Zur Elektrodynamic der bewegter K\"orper}, Annalen der Physik {\bf XVII}
(1905), 891--921.}

\bib{3}{J.\ Ehlers, F.\ A.\ E.\ Pirani, A.\ Schild, {\sl The Geometry of Free Fall and the Light
Propagation}, in ``Studies in Relativity: Papers in honour of J.\ L.\ Synge'' (1972), 63--84.}

\bib{4}{T.\ Levi-Civita, {\sl The Absolute Differential Calculus}, Blackie and Son (1929).}

\bib{5}{M.~Ferraris, M.~Francaviglia, C.~Reina, {\sl Variational Formulation of General Relativity
from 1915 to 1925 ``Palatini's Method'' Discovered by Einstein in 1925\/} - Gen.\ Rel.\ Grav.\ {\bf 14}
(1982) 243-254.}

\bib{6}{Hermann Weyl, Ann.~Phys.\ {\bf 59} (1919), 101.}

\bib{7}{S.\ Capozziello, M.\ Francaviglia, {\sl Extended Theories of Gravitation and Their
Cosmological and Astrophysical Applications}, Gen.\ Rel.\ Grav.\ {\bf 40}, 2-3, (2008), 357--420.}

\bib{8}{G.\ Magnano, M.\ Ferraris, M.\ Francaviglia,
{\sl Legendre Transformation and Dynamical Structure of Higher-Derivative Gravity}, Class.\ Quantum
Grav.\ {\bf 7} (1990) 557-570.}

\bib{9}{G.\ Magnano, M.\ Ferraris, M.\ Francaviglia,
{\sl Nonlinear Gravitational Lagrangians}, Gen.\ Rel.\ Grav.\ {\bf 19}(5) (1987) 465--479.}

\bib{10}{A.\ Eddington, {\sl The Mathematical Theory of Relativity}, Cambridge University Press
(1923).}

\bib{11}{J.\ Kijowski, {\sl On a purely affine formulation of general relativity}, Part II
Proceedings of the Conference Held at Salamanca September 10-Ð14 (1979), Edited By P.\ L.\ Garc\'ia,
A.\ P\'erez-Rend\'on.}

\bib{12}
 {G.~Allemandi, M.~Francaviglia, {\sl The Variational Approach to Alternative Theories of Gravity},
  KARP-2006-02, 2007. Prepared for 42nd Karpacz Winter School of Theoretical Physics: Current 
  Mathematical Topics in Gravitation and Cosmology, Ladek, Poland, 6-11 Feb 2006. Published in
  eConf C0602061:02, 2006; Int.~J.~Geom.~Meth.~Mod.~Phys.~4:1-23, 2007.}

\bib{13}
     {M.~Ferraris, M.~Francaviglia, I.~Volovich, {\sl  Universality of Einstein Equations},
      Class.\ Quantum Grav.\ {\bf 11} {\bf 6} (1994), 1505--1517, gr-qc/9303007.}

\bib{14}{T.\ P.\ Sotiriou, V.\ Faraoni, {\sl f(R) Theories of Gravity}, (2008), arXiv:0805.1726.}

\bib{15}
     {A.\ Pais, {\sl Subtle is the Lord: The Science and the Life of Albert Einstein}, Oxford
      University Press (1983).}

\bib{16}{S.\ Capozziello, V.F. \ Cardone, A. \ Troisi, {\sl Dark Energy and Dark Matter as curvature
effects?} JCAP {\bf 08}  (2006) 001--14.}

\bib{17}
 {J.-M.~Alimi, A.~F\"uzfa, {\sl The Abnormally Weighting Energy Hypothesis: the Missing Link between
 Dark Matter and Dark Energy}, arXiv:0804.4100.}

%==================================================================================================%
\bye